# Effect of spin transfer torque on the magnetic domain wall ferromagnetic resonance frequency in the nanowires

Jungbum Yoon[1], Chun-Yeol You[1], Younghun Jo[2], Seung-Young Park[2], and Myung-Hwa Jung[3]

[1]Department of Physics, Inha University, Incheon 402-751, Korea

[2]Division of Materials Science, Korea Basic Science Institute, Daejeon 305-333, Korea

[3]Department of Physics, Sogang University, Seoul 121-742, Korea

We investigate the influence of the domain wall ferromagnetic resonance frequency on the spin transfer torque in a ferromagnetic nanowire. By employing micromagnetic simulations with the spin transfer torque, we find that the domain wall resonance frequency decreases with increasing spin polarized current density, when there is no change in the resonance frequency of the domain itself. Surprisingly, the variation of the resonance frequency is remarkable (> 1.6 GHz) with the spin transfer torque even though the domain wall is pinned. Since the presented domain wall ferromagnetic resonance study has been performed for the pinned domain wall, the contributions of extrinsic defects are excluded. It is strong advantages of the present study, since the effects of extrinsic pinning sites are inevitable in the imaging or transport measurements.

The spin transfer torque (STT) is the phenomenon that the angular momenta are transferred from the spin-polarized conduction electrons to the localized spins in ferromagnets.[1,2] Recently, the novel spintronic devices with the STT are actively investigated such as the magnetic race-track memory,[3,4,5] spin-transfer torque magnetoresistive random access memory,[6,7,8] and spin torque nano oscillator.[9,10] The magnetic race-track memory based on the current-induced domain wall motion (CIDWM) in the ferromagnetic nanowires is one of the candidates for the next generation non-volatile mobile memories. The STT in the nanowire is important not only for the race-track memory applications, where the domain wall (DW) is moved by the STT, but also it has been revealed that many interesting physics are involved in the DW motion.[4] Since the DW width of the 3$d$ transition metal ferromagnetic nanowire is much larger than other relevant physical length scales such as Fermi wavelength and exchange length, the non-adiabatic contribution has been expected to be small.[11] However, many experimental observations have been reported that the non-adiabatic contribution is not small and it is important[12,13,14] even for the Permalloy. Furthermore, there are many controversial theories about magnitude of non-adiabatic contributions.[15,16,17] Many experimental efforts have been determined the detail interactions between conducting spins and localized spins inside of the DW.[18,19,20,21,22,23,24,25] Most experiments can be categorized by two kinds: one is transport measurements[20-23] and the other is magnetic domain imaging.[24-27] The transport measurements are able to determine whether the DW placed between two electrodes or not by magneto-resistance. Since it is relatively simple and easy to repeat, this method is widely used. However, the transport data cannot determine real DW position in the single ferromagnetic nanowire, and it requires many electrodes.[18,19] The monitoring DW motion by MOKE (magneto-optical Kerr effect) at the fixed point is relatively simple but it gives limited information whether the DW passed at specific position or not.[25] Direct imaging by magnetic force microscope is only possible to determine the static magnetization configuration.[22] And x-ray imaging can measure the dynamics of the DW with an assumption of repeated motion of the DW.[24] It must be emphasized that none of them are extracted any physical information without DW movement. However, the DW movement is strongly affected by extrinsic pinning sites. Therefore, any physical quantities extracted from the experimental measurements always include the extrinsic effect.

In this study, we would like to introduce the new possibility in the study of STT effect in the nanowires, DW-FMR (domain wall ferromagnetic resonance). Magnetic thermal noise, ferromagnetic resonance (FMR), and spin diode effects are widely accepted excellent experimental tools for the study of STT in magnetic tunneling junction geometry.[8,26,27,28] However, such methods have been applied to only limited study for the DW. According to our previous study, it has been revealed that the DW-FMR frequency is distinguishable from one of domain body itself due to the different effective field in the DW inside.[29] It can be easily understood that when spin polarized currents are applied to the nanowire, the spin dynamics of DW will be altered by the STT. The STT acts as an additional torque for each spin, even though the DW is pinned. This is the key idea of this study. We investigate the DW-FMR spectra of the non-moving DW by employing micromagnetic simulations including STT effect.[30] Even though when the DW distortion is too small to notice with STT, surprisingly, the variation of the resonance frequency is remarkable. Consequently, we can exclude the contributions of extrinsic pinning sites which govern the DW movement.

In order to investigate the effect of the spin polarized current on the spin dynamics of the DW, we employed the Landau-Lifshitz-Gilbert (LLG) equation with the STT contribution.[30]

$$d\vec{m}/dt = -|\gamma|\vec{H}_{eff} \times \vec{m} + \alpha \vec{m} \times d\vec{m}/dt - (\vec{u} \cdot \nabla)\vec{m} + \beta \vec{m} \times \left[(\vec{u} \cdot \nabla)\vec{m}\right]. \qquad (1)$$

Here $\gamma$, $\alpha$, $M_s$, $\vec{H}_{eff}$, and $\vec{m}$ are the gyromagnetic ratio, Gilbert damping constant, saturation magnetization, effective magnetic field, and unit vector of the magnetization, respectively. In Eq. (1), the last term with the dimensionless parameter $\beta$, compared to $\alpha$, presents the non-adiabatic spin torque. The x-direction is the longitudinal direction of the ferromagnetic nanowire as shown in Fig. 1. The velocity $u$ is defined parallel to the direction of the conductive electron, +x-direction, with the amplitude of $u = JPg\mu_B/2eM_s$, where $J$ is the current density and $P$ is the spin polarization rate. The majority spin direction is right for $u > 0$. $g\mu_B/2eM_s$ is given by $7 \times 10^{-11}$ m$^3$/C for Permalloy (Py).[16] And the material parameters of Py used in our simulation are as follows: the saturation magnetization $M_s = 8.6 \times 10^5$ A/m, the exchange stiffness $A_{ex} = 13 \times 10^{-12}$ J/m, the gyromagnetic ratio $\gamma = 2.21 \times 10^5$ m/(A·s), and we ignore the magnetocrystalline anisotropy. In this simulation, the Gilbert damping parameter $\alpha = 0.02$ is fixed. We perform our simulation at a zero temperature and take a cell size of $5 \times 5 \times 5$ nm$^3$.

In order to pin the DW, we introduce a small notch ($5 \times 10 \times 10$ nm$^3$) at the center of the 10-nm thick, 80-nm wide, and 2000-nm long Py nanowire as shown in Fig. 1. Initially, two magnetic domains are set up at each side of the nanowire without any external magnetic field and current. A stable tail-to-tail transverse type DW is formed with the energy minimization, and the stable DW is used as an initial magnetization configuration.[31] It must be noted that our study is limited for the transverse type DW.

In order to mimic FMR in our micromagnetic simulations, a "sine cardinal (sinc)" function $H_y(t) = H_0 \sin[2\pi f_H(t-t_0)]/2\pi f_H(t-t_0)$ is applied with $H_0 = 10$ mT and the field frequency $f_H = 45$ GHz to the whole nanowire area.[32] Even though $H_0$ is smaller, the results of FMR spectra are the same with smaller amplitude. Since the sinc function is a Fourier transform of $H_y(f) = H_0$ ($|f| < f_H$), the Fourier transform of the response of $H_y(t)$ is the FMR spectra of corresponding frequency. The time varying transverse magnetization component $M_y(x, y, t)$ configurations are stored in the whole Py nanowire at each temporal moment ($10^{-2}$ ns step) with the simulation time per $10^{-5}$ ns step. The results of the simulation are stored with each point for sufficiently long time (100 ns). The FMR spectra due to the RF-magnetic field are obtained by the fast Fourier transform (FFT) of $M_y(x, y, t)$.

In Fig. 2, we plot the typical FMR spectra of the Py nanowire, the FFT of $\sum_{x,y \in NW} M_y(x,y,t)$, without external dc-field and current. Here NW means a whole nanowire area. Since there is no external dc-field, the resonance frequencies of two opposite domain are the same, but there is clear additional resonance peak at 7.1 GHz (red arrow). According to the simple Kittel's equation, $f = (\gamma/2\pi) \times \sqrt{(H_{eff} + (N_z - N_x) \times M_s) \times (H_{eff} + (N_y - N_x) \times M_s)}$, (The $N_x$, $N_y$, and $N_z$ are the demagnetization factors), we can identify the 10.0 GHz peak (blue arrow) is corresponding to the resonance peak of the two domains. We also confirm that the peak at the 7.1 GHz is originated from the DW.[29] To clarify the DW contribution, the local magnetization, $\sum_{x,y \in DW} M_y(x,y,t)$, where DW means the region around the DW, are only considered as shown in Fig. 1 with a black rectangle. More details of local spectra concept already reported in our previous report.[29] As a result of the local spectra analysis, we find out that the DW-FMR frequency is lower than that of the domain. In the inset of Fig. 2, the local FMR spectra are plotted for the only DW region. Because the direction of each spin inside of the DW is gradually tilted from the one direction to the opposite one, the effective fields of the DW are non-uniform and the demagnetization field of the DW is smaller than the one of the domain. As a result, the resonance frequency decreases with smaller effective field. Even though, we cannot find analytic expression for the effective field of the inside of the DW, but it is clear that the inside effective field of the DW is quite different (smaller) from one of the domain.

In order to reveal the effect of STT in DW-FMR spectra, the velocity $u$ representing the spin polarized current density is applied from 0 to 400 m/s with 100 m/s steps for the Py nanowire with $\beta = 0.01$. First, we apply the spin polarized current and wait for 100 ns till the DW is stabilized. Due to the STT, the DW is slightly deformed, but it is hard to notice with DW images as shown in Fig. 3 (a)~(e). It must be emphasized that all the DWs in our study are almost identical shapes. Therefore, it is impossible to distinguish them by the magnetic imaging or transport measurement experiments. While the spin polarized current is keeping, the FMR simulations are performed. The local spectra for the DW are depicted in Fig. 4 (a). In these spectra, two groups of peaks are found. There are other peaks around 2 GHz (blue circle). We identify that the STT induced whole DW vibration is the source of 2 GHz peaks. These peaks are also found with the STT only without RF-magnetic field. Therefore, we will pay our attention to the peaks around 7 GHz which are indicated by arrows in Fig. 4 (a). Figure 4 (b) shows the dependence of the DW-FMR frequency on the $u$. It is clearly shown that the DW-FMR frequency decreases noticeably with increasing $u$ (from 7.2 to 5.6 GHz). When we consider the distortion of the DW structure is very small, the change of the resonance frequency is dramatic. This is the central results of our study. It must be noted the DW is depinning when $u > 410$ m/s.

Now, let us consider the effect of a non-adiabatic contribution with various $\beta$ in DW-FMR spectra. The local spectra for various $\beta$ (= 0, 0.01, 0.02, and 0.04) are depicted in Fig. 5 with constant $u$ (= 100 m/s). The dependence of spectra on $\beta$ is very small, and it can be recognized only enlarged plot in the inset of Fig. 5. It implies that the contribution of non-adiabatic STT term to the DW-FMR spectra is small, and practically undetectable by experimentally.

Let us discuss about our findings. When the STT is applied, we find the DW-FMR frequency decreases. Since there is no analytic expression of susceptibility of the DW with the STT, further detail analysis is unavailable in current stage. However, it is clear that the adiabatic STT, $-(\vec{u}\cdot\nabla)\vec{m}$, acts an additional torque and it belongs to the nanowire ($xy$) plane, and its role is somewhat similar to the external magnetic field.[16] It gives translational motion of the DW as the external field does when the damping is ignored or the damping is canceled with a non-adiabatic term.[33] With the STT, the DW is tilted from a nanowire plane with an angle $\phi$, which is proportional to the $u$.[16] When the DW is tilted, the effective field is reduced. Therefore, the DW-FMR frequency decreases. It must be note that the frequency shifts are independent on the sign of the $u$ (not shown here). The tilting angle $\phi$ is proportional to the $u$, but the reduction of the effective field is related with $|\phi|$ so that the change of the DW-FMR frequency depends on $|u|$. One more observation is that the dramatic reduction of an amplitude and area of the peaks, which are related with the magnon excitations. Smaller amplitude or area of the peak indicates smaller excitation of magnon. In this stage, the physical origin of the suppression of the magnon excitation is not clear, but it is sure that it is quite different from the magnetic tunneling junction cases.[6-8]

When the STT is applied to the DW, the Gilbert damping term tilts the DW to the out-of-plane, and the demagnetization energy prevent the tilting, and the DW will stop. The role of the non-adiabatic STT is suppression of Gilbert damping, so that DW will move continuously. Therefore, the contribution of $\beta$ in the DW spectra is clear. It reduces the Gilbert damping, which is related the line-width of the peak. However, the line-width of the DW-FMR spectra depends on not only the Gilbert damping but also the DW width. The non-uniform magnetization direction in the DW causes extra broadening of the line-width in the DW-FMR spectra. According to our micromagnetic simulations, the effect of $\beta$ is very small as shown in Fig. 5. Therefore, unfortunately, our proposed method will extract the only adiabatic information of the STT.

Finally, we would like to discuss about the experimental setup. Since the portion of the DW area is much smaller than the domain area, the FMR signal is much smaller than one of the whole sample. Therefore, an extra caution is required in the design of sample and electrodes. Or, BLS (Brillouin Light Scattering) microscopes might be proper tools.[34,35] Furthermore, the current introduces inevitable Joule heating,[36,37] and the higher temperature causes red shift of the resonance frequency, which is the same direction of adiabatic STT contribution. Therefore, a separate resistance measurement to monitoring the temperature of the sample must be considered.

In conclusion, we proposed the DW-FMR spectra measurement in the study of the STT in the nanowire. We found that the DW-FMR frequency is very sensitive on the adiabatic STT, even the DW shape is barely changed. It must be emphasized that the presented DW-FMR is free from the contribution of extrinsic pinning sites, which is inevitable in other measurement methods.

Acknowledgement

This work was supported by Nano R&D program (Grant No. 2008-02553), by KBSI grant (T30405) for Y. Jo, by the IT R&D program of MKE/KEIT [2009-F-004-01], and by the Sogang University Research Grant of 201011014.01.

Figure Captions

Fig. 1 Magnetization configuration of the Py nanowire with a transverse Néel type DW. The DW pinned at the notch in a center of the nanowire.

Fig. 2 FMR spectra of the Py nanowire with $u = 0$ m/s. The red and blue arrows indicate the resonance frequencies of the domain wall and domain, respectively. Inset represents the local DW-FMR spectra.

Fig. 3 (a)-(e) Magnetization configuration of the DW with various $u = (0 \sim 400$ m/s$)$ with 100 m/s steps. The figures are enlarged on the DW with the arrows of macro-magnet directions.

Fig. 4 (a) Local DW-FMR spectra for various $u$. The blue circles indicate the vibration of the DW. (b) Dependence of the DW-FMR frequency on $u$. The velocity $u$ is proportional to the current density.

Fig. 5 DW-FMR spectra for various non-adiabatic STT $\beta$ with $u = 100$ m/s. Inset represents the enlargement of the resonance peaks.

References

1  J. C. Slonczewski, J. Magn. Magn. Mater. **159**, L1 (1996).
2  L. Berger, Phys. Rev. B **54**, 9353 (1996).
3  S. S. P. Parkin, U.S. Patent No. 6834005, (2004).
4  S. S. P. Parkin, M. Hayashi, and L. Thomas, Science **320**, 190 (2008)
5  M. Hayashi, L. Thomas, R. Moriya, C. Rettner, and S. S. P. Parkin, Science **320**, 209 (2008).
6  S.-C. Oh, S.-Y. Park, A. Manchon, M. Chshiev, J.-H. Han, H.-W. Lee, J.-E. Lee, K.-T. Nam, Y. Jo, Y.-C. Kong, B. Dieny, and K.-J. Lee, Nature Phys. **5**, 898 (2009).
7  C.-Y. You, Curr. Appl. Phys. **10**, 952 (2010).
8  M. H. Jung, S. Park, C.-Y. You, and S. Yuasa, Phys. Rev. B **81**, 134419 (2010).
9  A. Bisig, L. Heyne, O. Boulle, and M. Kläui, Appl. Phys. Lett. **95**, 162504 (2009).
10 C.-Y. You, J. Magnetics **14**(4) 168 (2009).
11 J. Xiao, A. Zangwill, and M. D. Stiles, Phys. Rev. B **73,** 054428 (2006).
12 G. S. D. Beach, C. Knutson, C. Nistor, M. Tsoi, and J. L. Erskine, Phys. Rev. Lett. **97**, 057203 (2006).
13 S. Yang and J. L. Erskine, Phys. Rev. B **75**, 220403(R) (2007).
14 S. Lepadatu, A. Vanhaverbeke, D. Atkinson, R. Allenspach, and C. H. Marrows, Phys. Rev. Lett. **102**, 127103 (2009).
15 S. Zhang and Z. Li, Phys. Rev. Lett. **93**, 127204-1 (2004).
16 A. Thiaville, Y. Nakatani, J. Miltat and Y. Suzuki, Europhys. Lett. **69**, 990 (2005).
17 J. He, Z. Li, and S. Zhang, J. Appl. Phys. **98**, 016108 (2005).
18 M. Kläui, C. A. F. Vaz, J. A. C. Bland, W. Wernsdorfer, G. Faini and E. Cambril, Appl. Phys. Lett. **81**, 108 (2002).
19 M. Kläui, C. A. F. Vaz, J. A. C. Bland, W. Wernsdorfer, G. Faini, E. Cambril, L. J. Heyderman, F. Nolting, and U. Rüdiger, Phys. Rev. Lett. **94**, 106601 (2005).
20 M. Hayashi, L. Thomas, Y. B. Bazaliy, C. Rettner, R. Moriya, X. Jiang, and S. S. P. Parkin, Phys. Rev. Lett. **96**, 197207 (2006).
21 L. Thomas, M. Hayashi, X. Jiang, R. Moriya, C. Rettner, and S. S. P. Parkin, nature **443**, 197 (2006).
22 A. Yamaguchi, T. Ono, S. Nasu, K. Miyake, K. Mibu, and T. Shinjo, Phys. Rev. Lett. **92**, 077205-1 (2004).
23 M. Kläui, P.-O. Jubert, R. Allenspach, A. Bischof, J. A. C. Bland, G. Faini, U. Rüdiger, C. A. F. Vaz, L. Vila, and C. Vouille, Phys. Rev. Lett. **95**, 026601 (2005).
24 G. Meier, M. Bolte, R. Eiselt, B. Krüger, D.-H. Kim, and P. Fischer, Phys. Rev. Lett. **98**, 187202 (2007).
25 K.-J. Kim, J.-C. Lee, S.-M. Ahn, K.-S. Lee, C.-W. Lee, Y. J. Cho, S. Seo, K.-H. Shin, S.-B. Choe and H.-W. Lee, nature **458**, 740 (2009).
26 Y. Guan, D. W. Abraham, M. C. Gaidis, G. Hu, E. J. O'Sullivan, J. J. Nowak, P. L. Trouilloud, D. C. Worledge, and J. Z. Sun, J. Appl. Phys. **105**, 07D127 (2009).
27 S. Petit, N. de Mestier, C. Baraduc, C. Thirion, Y. Liu, M. Li, P. Wang, and B. Dieny, Phys. Rev. B **78**, 184420 (2008).
28 A. A. Tulapurkar, Y. Suzuki, A. Fukushima, H. Kubota, H. Maehara, K. Tsunekawa, D. D. Djayaprawira, N. Watanabe, and S. Yuasa, nature **438**, 339 (2005).
29 J. Yoon, C.-Y. You, Y. Jo, S.-Y. Park, and M.-H. Jung, cond-mat/1006.5763.
30 See http://www.zurich.ibm.com/st/magnetism/spintevolve.html
31 Y. Nakatani, A. Thiaville, and J. Miltat, J. Magn. Magn. Mater. **290–291**, 750 (2005).
32 K.-S. Lee, D.-S. Han, and S.-K. Kim, Phys. Rev. Lett. **102**, 127202 (2009).
33 M. D. Stiles, W. M. Saslow, M. J. Donahue, and A. Zangwill, Phys. Rev. B **75**, 214423 (2007).
34 C. Bayer, J. Jorzick, B. Hillebrands, S. O. Demokritov, R. Kouba, R. Bozinoski, A. N. Slavin, K. Y. Guslienko, D. V. Berkov, N. L. Gorn, and M. P. Kostylev, Phys. Rev. B **72**, 064427 (2005).
35 H. Schultheiss, S. Schäfer, P. Candeloro, B. Leven, B. Hillebrands, and A. N. Slavin, Phys. Rev. Lett. **100**, 047204 (2008).
36 C.-Y. You, I. M. Sung, and B.-K. Joe, Appl. Phys. Lett. **89**, 222513 (2006).
37 C.-Y. You and S.-S. Ha, Appl. Phys. Lett. **91**, 022507 (2007).

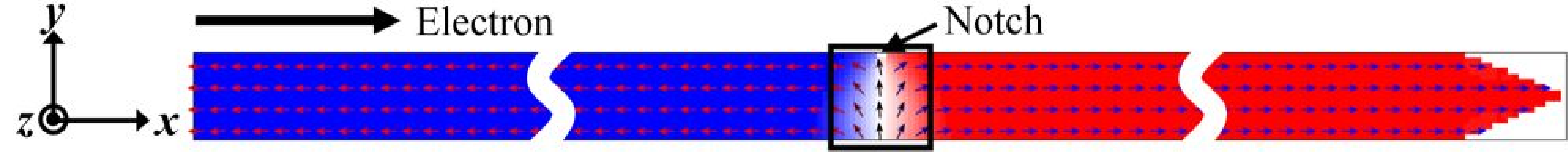

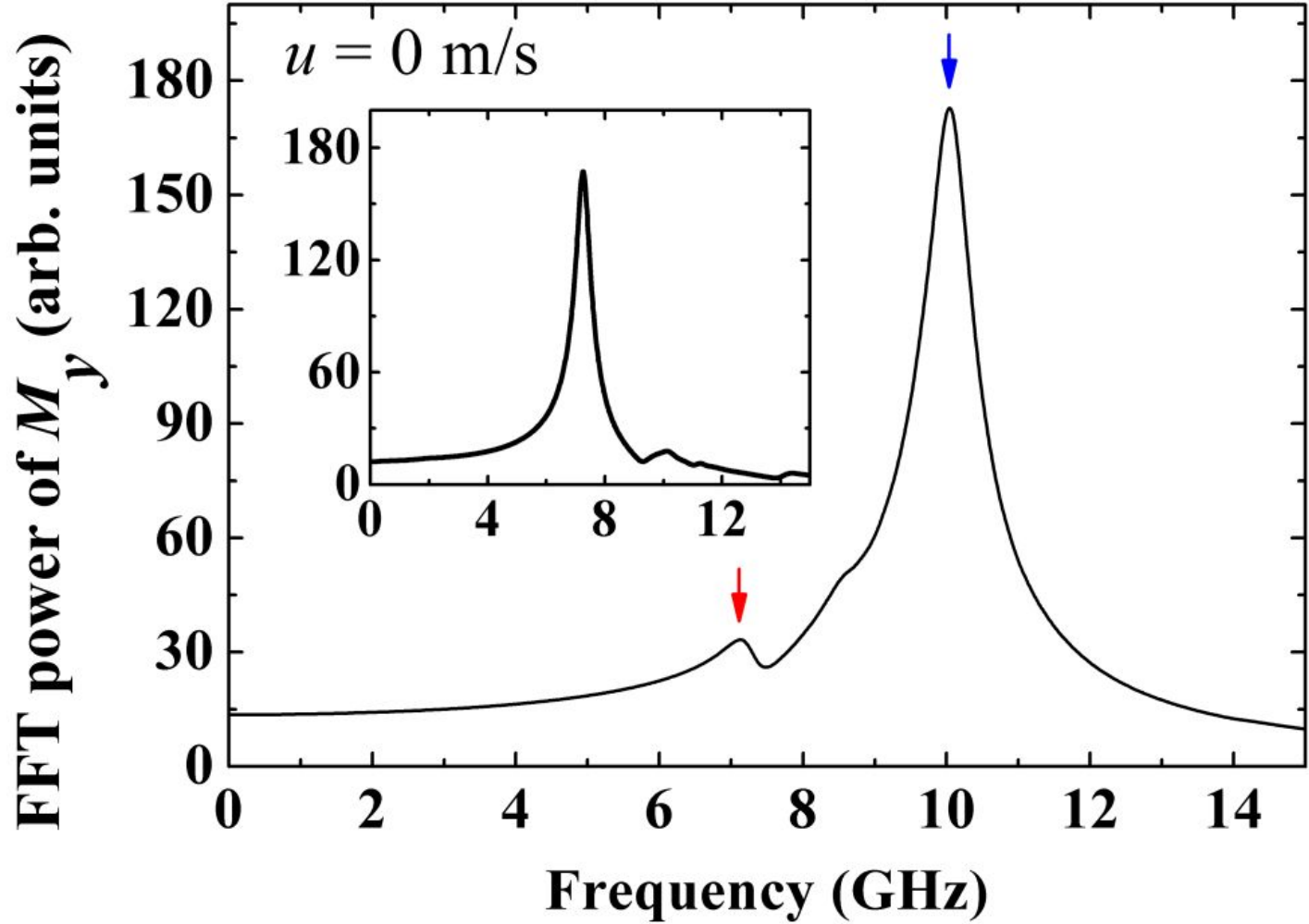

(a) $u = 0$  (b) 100  (c) 200  (d) 300  (e) 400

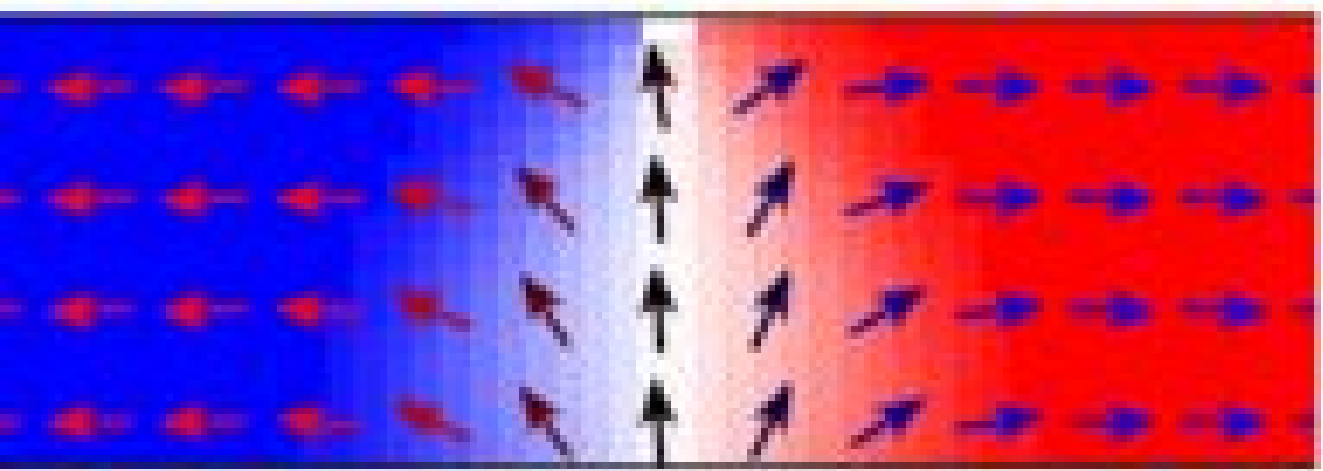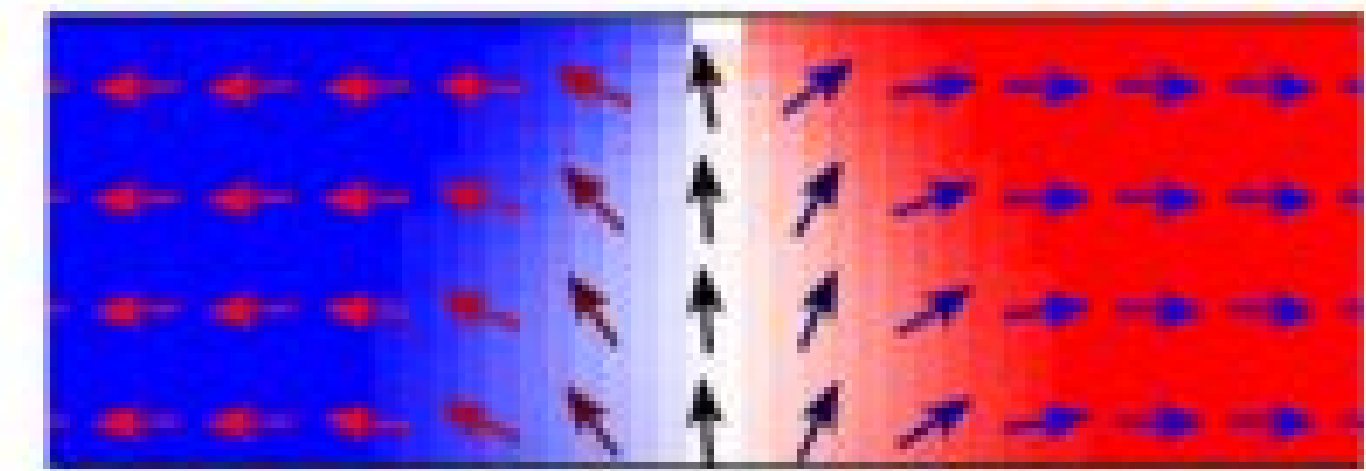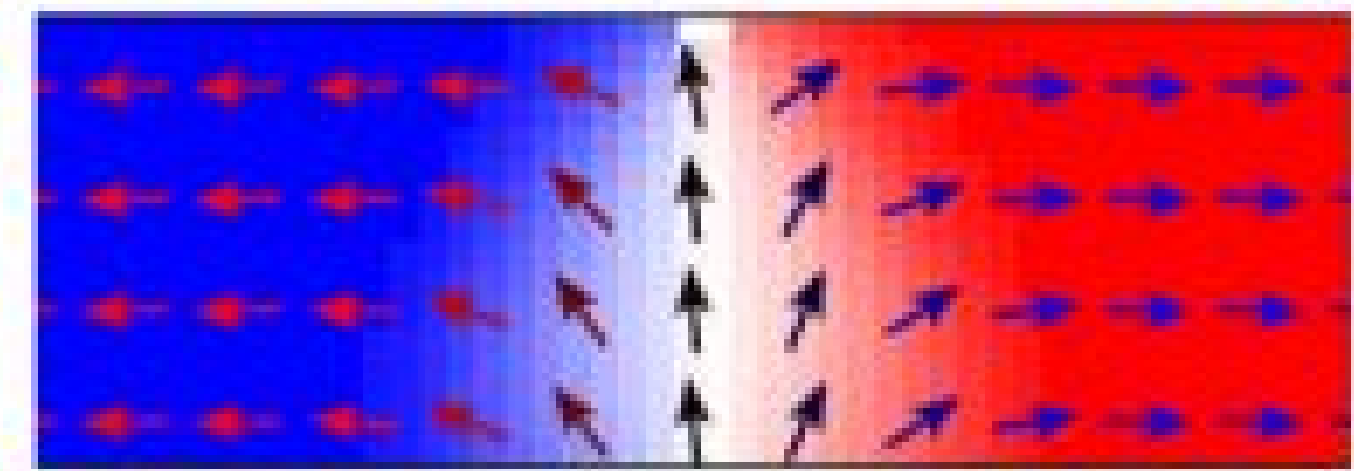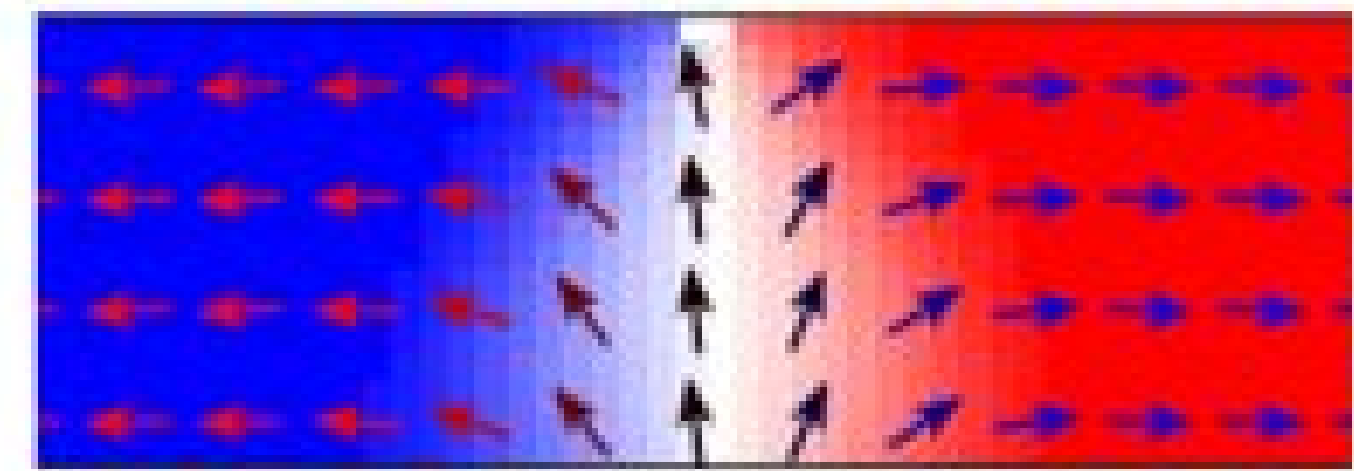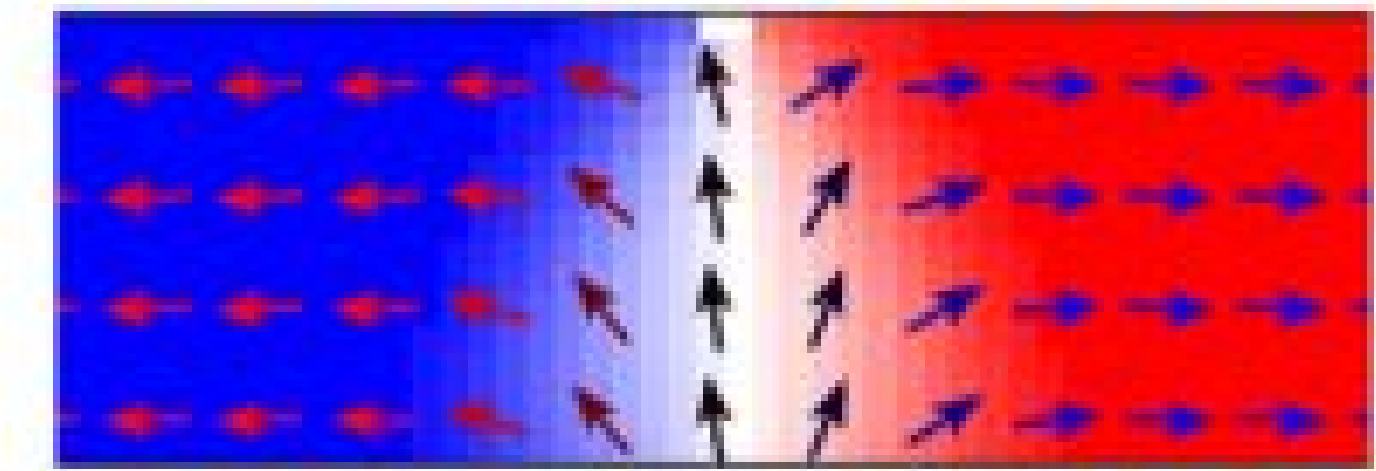

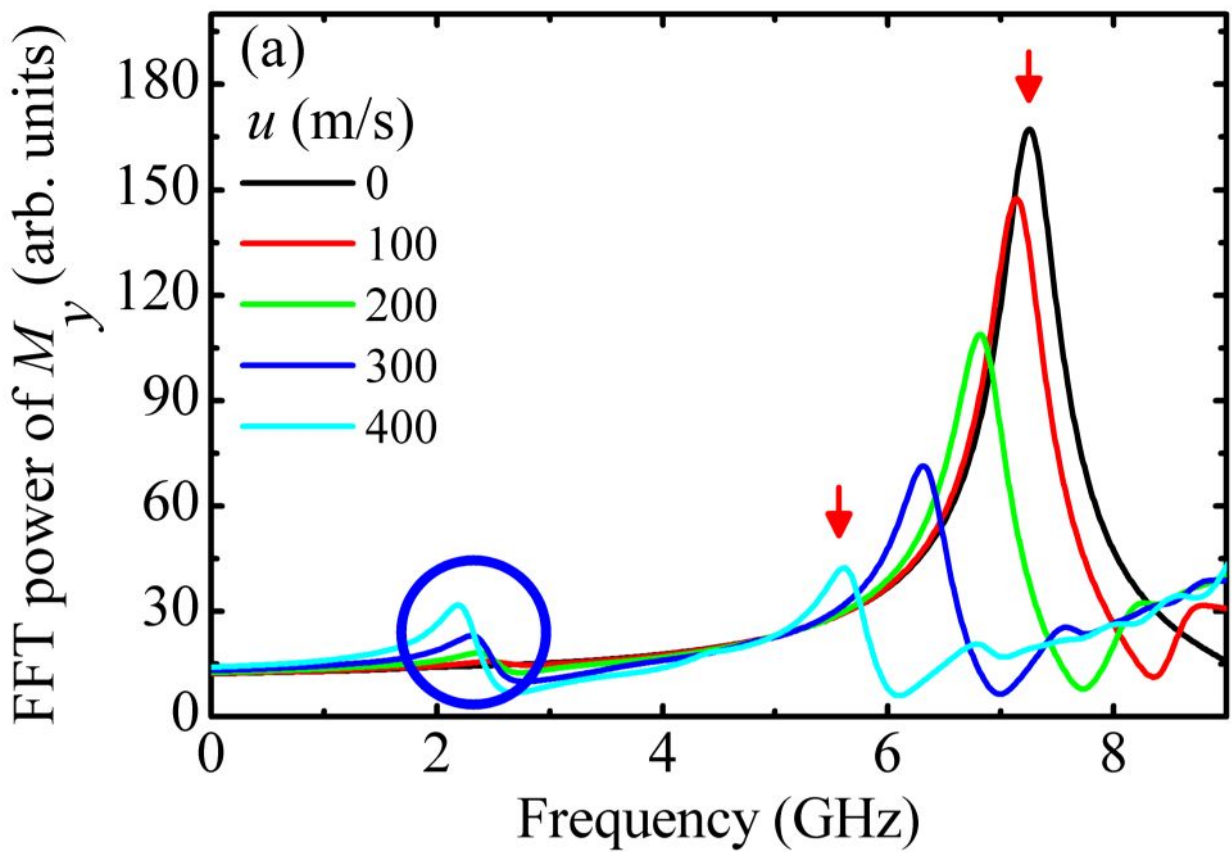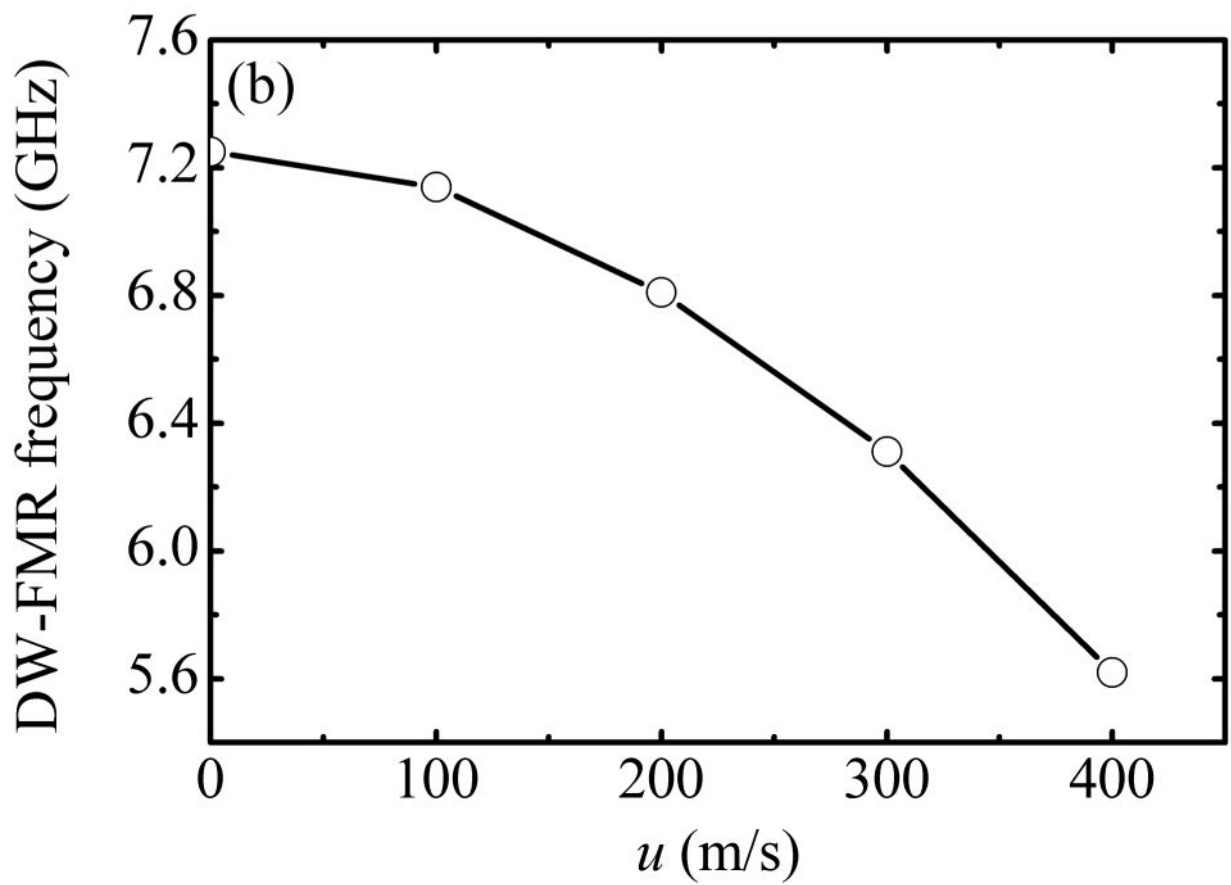

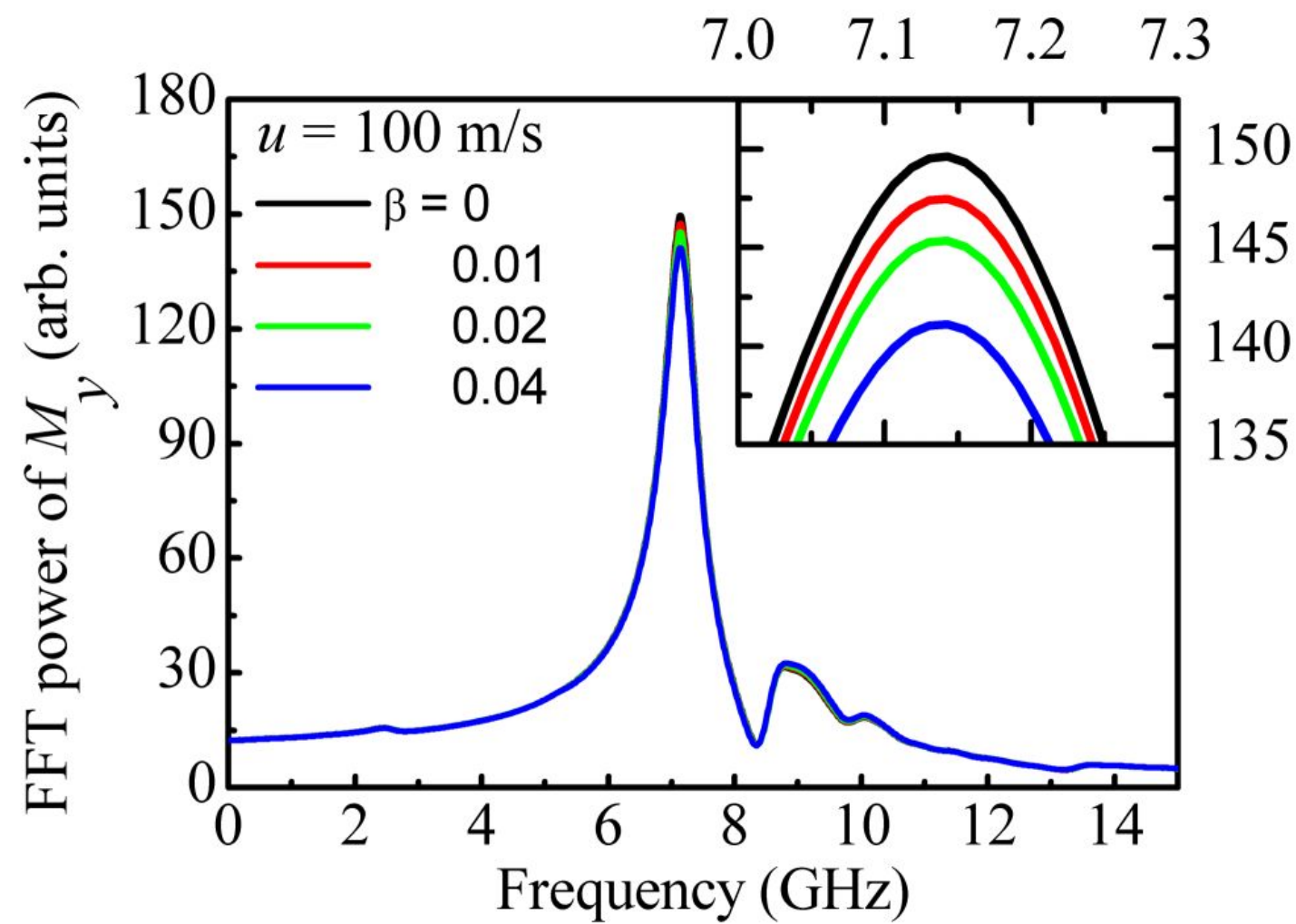